# Resolving Rydberg-Electron Recapture Dynamics via Laser-driven Frustrated Tunneling Ionization

Sainan Peng[1,2,§], Yudong Chen[1,2,§], Yang Li[3,†], Guangyu Fan[4], Xinhua Xie[5], Feng He[3], Zhensheng Tao[1,2,*]

[1]*State Key Laboratory of Surface Physics and Key Laboratory of Micro and Nano Photonic Structures (MOE), Department of Physics, Fudan University, Shanghai 200433, P R China.*
[2]*Shanghai Key Laboratory of Metasurfaces for Light Manipulation, Fudan University, Shanghai 200433, P R China.*
[3]*Key Laboratory for Laser Plasmas (Ministry of Education) and School of Physics and Astronomy, Collaborative Innovation Center for IFSA (CICIFSA), Shanghai Jiao Tong University, Shanghai 200240, P R China.*
[4]*Shanghai Key Lab of Modern Optical System, University of Shanghai for Science and Technology, Shanghai, 200093, China.*
[5]*SwissFEL, Paul Scherrer Institute, Villigen PSI 5232, Switzerland.*
[§]These authors contributed equally to this work.

**ABSTRACT**:

By employing two-color counter-rotating circularly polarized laser fields, we investigate the dynamics of electron recapture into Rydberg states under strong, ultrashort laser pulses, probed via coherent extreme-ultraviolet free-induction decay (XFID). Our study reveals significant distinctions between XFID and above-threshold high-order harmonic generation in terms of their ellipticity dependence on the driving-laser waveforms, yield variations with the laser-intensity ratios, and sensitivity to the driving-laser ellipticity. All these differences arise from the fundamentally distinct electron trajectories underlying the two processes. More importantly, our findings provide compelling evidence that Rydberg-electron recapture predominantly occurs at the end of the driving laser field, offering the first direct experimental confirmation of this long-proposed mechanism.

---

[†] liyang22@sjtu.edu.cn

[*] zhenshengtao@fudan.edu.cn

*Introduction.* – Ionization plays a crucial role in the study of laser-matter interactions. With advancements in laser technologies, our understanding of ionization has significantly expanded. Starting from Einstein's photoelectric effect, the field has progressed through various ionization mechanisms, including multiphoton ionization, tunneling ionization, ionization stabilization, rescattering double ionization, and other fascinating phenomena [1–4]. In addition to producing photoelectrons, which carry abundant information about the target, strong-field ionization also leads to high-order harmonic generation (HHG) [5–8], which has attracted considerable attention for its wide applications, such as synthesizing attosecond pulses [9–12].

Among these ionization mechanisms, frustrated tunneling ionization (FTI) was proposed to explain the formation of Rydberg states in strong laser fields. In FTI, an electron is initially liberated near the crest of the laser electric field and subsequently recaptured into high-lying Rydberg states [13–19]. Rydberg-state excitation through FTI can be harnessed for applications such as accelerating and decelerating neutral particles [20,21], generating narrow-linewidth extreme ultraviolet (XUV) radiation [22–25], and realizing multiphoton Rabi oscillations [26,27].

While the mechanism of laser-driven FTI has been extensively investigated, most studies have employed single-color driving lasers. By examining how the fraction of neutral excited atoms [17] or the intensity of XUV free-induction decay (XFID) [24,25] depends on laser parameters such as ellipticity and carrier-envelope phase, these studies have highlighted the coherence of FTI and its connection to the tunneling-plus-rescattering mechanism [17]. However, these studies have primarily focused on the resemblance between FTI and HHG [see Supplementary Materials (SM) Section S2], offering limited insights into the timing and dynamics of electron recapture into Rydberg states. While techniques such as attosecond light-house have demonstrated that Rydberg-electron recapture involves longer trajectories compared to HHG [24], the quantitative details of these trajectories and the recapture mechanisms remain elusive. Moreover, many theoretical models have assumed that Rydberg-electron recapture predominantly occurs at the end of the driving pulse [14,28,29], yet this hypothesis so far lacks experimental verification. The primary limitation of single-color driving lasers lies in their limited degrees of freedom, which hinder the flexible manipulation of electron trajectories and the precise timing of the recapture process.

In this work, we address these challenges by employing two-color counter-rotating circularly polarized (TCCP) driving fields to experimentally and theoretically investigate the recapture dynamics of electrons into Rydberg states under strong, ultrashort laser pulses, as probed through XFID radiations. TCCP fields offer exceptional flexibility in controlling three-dimensional electron trajectories, enabling the examination of Rydberg-state formation through key parameters, including XUV ellipticity, and the intensity- and polarization-dependence on the driving lasers. Different from previous studies, our results reveal significant differences between the XFID and HHG processes: (1) XFID ellipticity is more sensitive to the broken symmetry of the driving laser field, (2) XFID radiation exhibits a fixed optimal intensity ratio of the TCCP fields, in stark contrast to HHG, and (3) XFID is orders of magnitude more sensitive to the driving-laser ellipticity than HHG. All of these differences arise from the characteristic electron trajectories of XFID, which span multiple optical cycles, in comparison to the short, single-cycle trajectories of HHG. Furthermore, the observed optimal intensity ratio provides compelling evidence that Rydberg-electron recapture predominantly occurs at the end of the driving laser pulse, offering the first direct experimental confirmation of the long-proposed mechanism.

*XFID and HHG spectra generated using TCCP laser fields.* – Figure 1a illustrates the concept of our experiments. The TCCP laser fields are composed of a near-infrared fundamental laser ($\omega$, 1030 nm, 34 fs) and its second harmonic ($2\omega$, 515 nm, ~75 fs) of opposite helicities. The intensities of both fields ($I_\omega$ and $I_{2\omega}$) can be independently adjusted by half-waveplate-polarizer pairs in their beam paths. XFID and HHG radiations are generated when the two driving pulses spatially and temporally overlap in a gas cell filled with argon gas. The resulting spectra are recorded by an XUV spectrometer. More details about the experimental setup can be found in SM Section S1.

Distinct XUV spectral features are observed below and above the ionization threshold (Figs. 1b and c). The HHG spectrum shows strong suppression of the $3n^{th}$ order harmonics (where $n$ is an integer), due to the three-fold symmetry of the driving fields, consistent with previous studies [30–33]. The $(3n-1)^{th}$ and $(3n+1)^{th}$ harmonics are circularly polarized with opposite helicities. The typical linewidth of these harmonics is ~0.17 eV (Fig. 1c). In contrast, the XUV spectrum below the threshold displays characteristic narrow spectral lines near the atomic resonances [24,25,34,35], with a linewidth of ~20 meV (Fig. 1b). These narrow XUV lines (R1, R2 and R3) are attributed to XFID, originating from the bound-bound transitions from the $ns$ and $nd$ states in the $Ar^+(^2P_{3/2})$ and $Ar^+(^2P_{1/2})$ manifolds [36] to the $p$-ground state [24,34].

The generation of XFID and HHG radiations driven by TCCP fields can be well reproduced by time-dependent Schrödinger equation (TDSE) simulations (Fig. 1d). In these simulations, a re-calibrated Tong-Lin type potential of argon is employed to closely reproduce the energies of the argon's excited states [37,38]. The contributions of the triple degenerate valence-shell *p* orbitals are coherently summed to obtain the XUV spectra. The XFID linewidths in our simulation are determined by the 80-fs time window employed for free propagation after the laser pulses (see SM Section S3). The TDSE simulations also elucidate distinct time-domain characteristics of the HHG and XFID radiations (Fig. 1e): HHG is generated through the tunnel ionization and sub-cycle recombination processes during the driving pulse, while XFID is strongly suppressed during the pulse and primarily emerges after the pulse has ended, emitted by electrons recaptured into the Rydberg states.

*Ellipticity dependence on the driving-laser waveforms.* – The ellipticities of XUV radiations were characterized using a reflective polarizer consisting of four gold mirrors, with $\theta_{\text{pol}}$ being the polarizer angle (Fig. 2a). Our results show that the TCCP-field-driven XFID can generate elliptically polarized XUV radiation, similar to HHG [30,31] (see insets of Fig. 2b). Notably, while the ellipticities of both radiations are insensitive to the intensity ratios $I_{2\omega}/I_{\omega}$, the XFID ellipticity ($\varepsilon_{\text{XFID}}\approx 0.6$) is consistently lower than that of HHG ($\varepsilon_{\text{HHG}}\approx 0.8$) under the same circular polarization of the TCCP fields (Fig. 2b).

The difference in ellipticities can be attributed to the contrasting electron trajectories involved in their generation. Notably, in our experiment, the full-width-at-half-maximum (FWHM) duration of the 515 nm pulse ($\tau_{2\omega}\approx$ 75 fs) is longer than that of the 1030 nm pulse ($\tau_{\omega}\approx$ 34 fs), which breaks the three-fold symmetry of the TCCP fields at the trailing edge of the envelope (Figs. 2c-e). Circularly polarized HHG is mainly generated by the electrons ionized at the center of the combined laser pulse ($t\approx 0$) that recollide with their parent ions on sub-cycle timescales under a nearly ideal trefoil driving field (Fig. 2d) [30,32,33]. At times far away from the center (e.g. $t\approx\pm30$ fs), the HHG emission is strongly suppressed due to both reduced ionization rates and non-closed electron trajectories (Fig. 2e). In contrast, for XFID, although it is also predominantly generated by electrons ionized close to $t\approx0$ [28], these electrons experience the entire trailing edge with broken three-fold symmetry before being recaptured (Fig. 2e), leading to reduced circular polarization.

Our explanation is supported by the TDSE simulations with varying pulse durations (Fig. 2b). When accounting for the difference in pulse durations ($\tau_{2\omega}\approx2\tau_\omega$, solid lines in Fig. 2b), the XFID ellipticity is generally lower than that of HHG, consistent with the experimental results. However, with the same driving laser polarizations but equal pulse durations ($\tau_{2\omega}=\tau_\omega$, blue dashed line in Fig. 2b), the XFID radiation exhibits a much higher ellipticity ($\varepsilon_{XFID}\approx0.9$). In addition, our simulations show that the HHG ellipticity remains nearly constant across different pulse durations, in stark contrast to XFID (see SM Section S4). Previous studies have shown that broken symmetry in TCCP fields could lead to a slight degradation of HHG circular polarization [39]. Here, our findings further demonstrate that the narrow-linewidth XUV radiation via XFID is much more sensitive to the driving-field symmetry, indicating multicycle electron trajectories involved.

*XUV yields as a function of the intensity ratios.* – In Figs. 3a and b, we present the experimental intensities the 22$^{nd}$-order HHG and of the XFID from the R1 Rydberg states, plotted as functions of the intensity ratio $I_{2\omega}/I_\omega$ under different total intensities $I_{\text{total}} = \left(\sqrt{I_\omega} + \sqrt{I_{2\omega}}\right)^2$. We find that the XFID intensity always peaks around a single $I_{2\omega}/I_\omega$ ratio of ~2.2 across a wide range of $I_{\text{total}}$ (Fig. 3b). In contrast, the optimal $I_{2\omega}/I_\omega$ ratios for HHG shift from 2.8 to 5.8 as $I_{\text{total}}$ increases from 1.5 to 2.9×10$^{14}$ W cm$^{-2}$ (Fig. 3a). This contrasting behavior is consistently observed across all the XFID (R1, R2 and R3) and HHG radiations (see SM Section S5). In Fig. 3c, we summarize the optimal intensity ratios ($\gamma_0$) for the two types of radiation.

To explore the underlying physics, we resort to classical-trajectory Monte Carlo (CTMC) simulations, which account for tunneling ionization and the classical evolution of liberated electrons under the combined influence of driving laser fields and Coulomb potential. For HHG, we examine electrons driven back to a distance from the parent ion smaller than their tunneling positions, with the corresponding electron energy equal to the harmonic energy plus the argon ionization potential. For XFID, electrons with final energies $E_f < 0$ can be recaptured into Rydberg states, and hence contribute to the radiation. More details about the CTMC simulations can be found in SM Section S6.

The measurement of the optimal intensity ratios, $\gamma_0$, enables precise timing of Rydberg-electron recapture under ultrashort laser pulses. Since only electrons with low final energies can be recaptured into Rydberg states, minimizing the energy acquired through laser acceleration is essential. Under TCCP fields, electrons ionized at three distinct field peaks within a single optical cycle share a common zero-vector-potential time ($t_{r0}$), where electrons have the highest probability of being recaptured. This recapture time is periodic, recurring every optical cycle of the fundamental field at $t_r = t_{r0} + NT_\omega$, where $N$ is a nonnegative integer and $T_\omega$ is the period of the fundamental driving field (see SM Section S6).

In Figs. 3d-f, we present the CTMC results for the recapture times corresponding to $N$=0, 2, and 9, respectively representing scenarios where electrons are recaptured after 1 cycle, 3 cycles, and 10 cycles of the fundamental field following the intensity peak of the driving laser pulses. For $N$=9, the recapture process occurs after the end of the driving pulse. Our results clearly demonstrate that the experimentally observed $\gamma_0 \approx 2.2$ can only be explained by long, multicycle electron trajectories, where electrons are recaptured after the end of the laser pulse. This conclusion is consistent with the TDSE results shown in Fig. 1e.

CTMC simulations also provide insights into the contrasting sensitivities of the optimal intensity ratio ($\gamma_0$) to the total laser intensity between XFID and HHG, as illustrated in Fig. 3c. In the semiclassical picture, the brightness of the XFID and HHG radiations is determined by the product of the ionization and recombination probabilities. The optimal $\gamma_0$ is mainly determined by the ionization probability, while the recombination probability increases monotonically with the ratio $I_{2\omega}/I_\omega$ (see SM Section S5). For HHG, the highest emission efficiency is achieved when the returning electrons collide head-on with the parent ions [40,41]. In contrast, for XFID, electrons are most efficiently recaptured into the Rydberg states when their momenta are near zero after the driving pulse [14,24]. These differences in the final states conversely determine different electron ionization times (birth time, $t_b$) of each process.

In Figs. 4a and b, we respectively present the electron distributions for XFID and HHG in the initial tunneling coordinates ($t_b$, $p_\perp$), where $p_\perp$ represents the initial transverse momentum (see SM Section S5). Unlike HHG, the recaptured XFID electrons exhibit a broad $t_b$ distribution. As $I_{total}$ varies from 1.5 to $3.0\times10^{14}$ W cm$^{-2}$, sub-cycle shifts in $t_b$ are observed in both cases. However, because of the broad $t_b$ distribution for XFID (Fig. 4c), the averaged field strength for tunneling ionization is not strongly affected, in contrast to HHG (Fig. 4d). This explains the insensitivity of $\gamma_0$ to $I_{total}$ for the XFID radiation (Fig. 3c).

*Sensitivity to the driving-laser ellipticity.* – Finally, the distinct laser-driven electron dynamics in XFID and HHG generation are further evidenced by their different responses to the driving-laser ellipticity. In this study, by maintaining the intensity ratio $I_{2\omega}/I_{\omega}$ at ~2.1, we simultaneously adjusted the ellipticities of the two-color driving lasers ($\varepsilon_\omega$ and $\varepsilon_{2\omega}$), while preserving the counter-rotating helicities. As shown in Fig. 4a, the HHG intensity generated by the circular polarized driving fields ($|\varepsilon_\omega| = |\varepsilon_{2\omega}| = 1.0$) decreases by only a factor of ~3 when compared to that driven by the linearly polarized fields ($\varepsilon_\omega = \varepsilon_{2\omega} = 0$) [30]. In stark contrast, the XFID intensity drops by two orders of magnitude. These results can also be well reproduced by the TDSE simulations (Fig. 4a).

The difference in ellipticity sensitivity stands in stark contrast to previous studies using single-color laser fields [24,25], arising from the distinct electron recapture trajectories in the two processes. In HHG, since ionized electrons can effectively recombine with their parent ions through closed trajectories, the generation efficiency is not greatly affected by the changes of the driving-laser polarizations. The observed variation in HHG efficiency primarily stems from the shifts in birth times and the consequent changes in ionization probability. In contrast, for XFID, low-energy electrons exhibit a higher probability of being captured by the Rydberg states. Since the electron's final energy after the laser pulse is proportional to the square of the vector potential at its birth time [$E_f \propto |A(t_b)|^2$], the minimum vector potential can become zero under linearly polarized driving fields (the red dot in Fig. 4b), whereas it remains a finite value under circular polarizations (the red dot in Fig. 4c). Consequently, the XFID intensity exhibits much greater sensitivity to the driving-laser polarization.

*Conclusion.* – We systematically compared the XFID and HHG radiations driven by TCCP laser fields, and uncovered key differences in ellipticity dependence, intensity-ratio dependence, and their sensitivity to laser ellipticity. All of these differences arise from the distinct electron trajectories involved in the two processes. Importantly, our results provide the first experimental confirmation of the long-proposed mechanism that Rydberg-electron recapture predominantly occurs after the end of the driving laser pulse. Furthermore, our findings demonstrate that the polarization states of coherent XFID can be controlled through light-wave engineering, paving the way for generating ultrashort, narrow-linewidth XUV radiations with structured polarization states in future applications [42–44].

**Acknowledgments**

We acknowledge financial support from the National Key Research and Development Program of China (Grant Nos. 2021YFA1400200 and 2022YFA1404700), the National Natural Science Foundation of China (Grant Nos. 12221004, 12274091, 12274294, 11925405, 12374318) and the Shanghai Municipal Science and Technology Basic

Research Project (Grant No. 22JC1400200, 19JC1410900). The computations in this paper were performed on the Siyuan-1 cluster, supported by the Center for High Performance Computing at Shanghai Jiao Tong University.

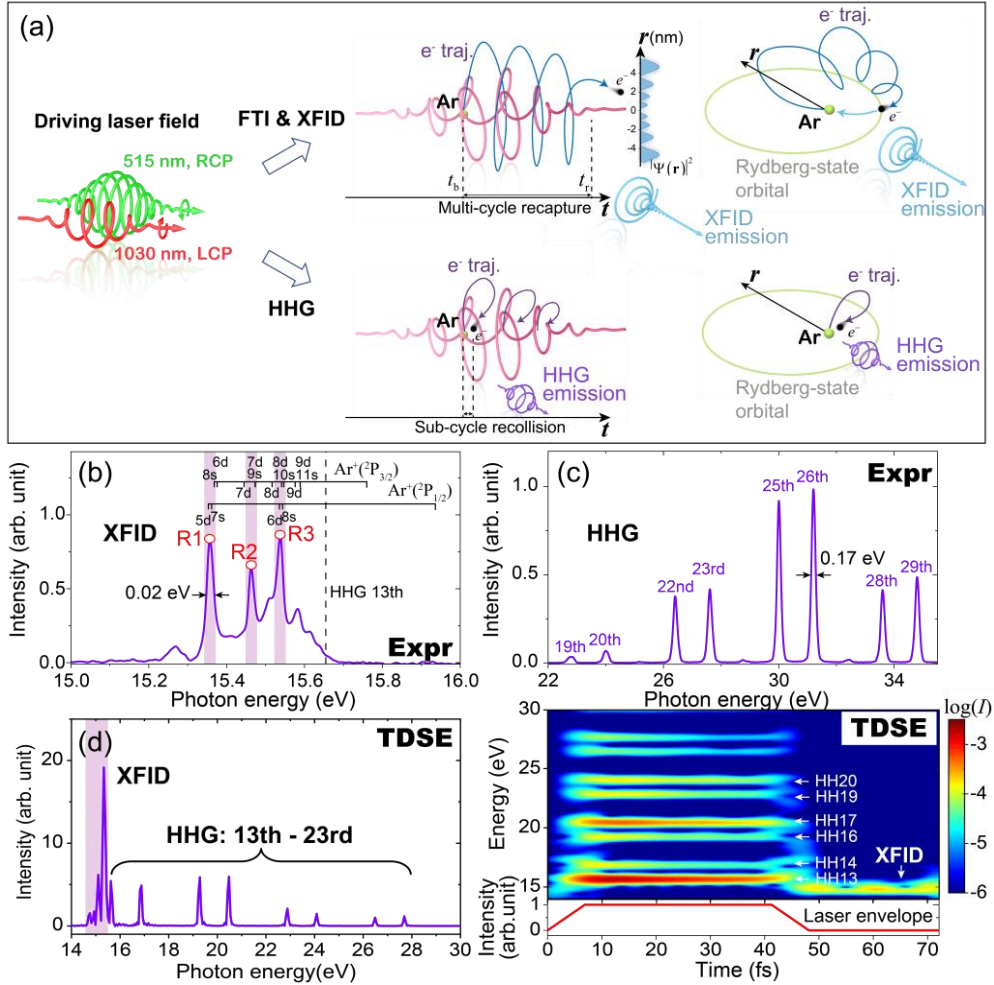

**Figure 1. (a)** Conceptual illustration: Both the XFID (and FTI) and HHG processes are driven by TCCP laser fields. XFID emission is generated via electrons recaptured by Rydberg states traversing multiple laser cycles from their birth time ($t_b$) to recapture time ($t_r$), while HHG emission results from sub-cycle recollision of ionized electrons. **(b)** Experimental XFID spectrum in argon with Rydberg manifolds shown for comparison. The energy of the 13$^{th}$-order harmonic is labeled by the dashed line. **(c)** Experimental HHG spectrum of argon. Typical linewidths are labeled. **(d)** XUV spectrum reproduced by the TDSE simulations, displaying both the XFID and HHG emissions. **(e)** Time-frequency analysis of the TDSE results: HHG emission occurs within the laser pulse, while XFID predominantly occurs after the pulse. **Lower panel:** Envelope of the driving laser pulse used in the simulation.

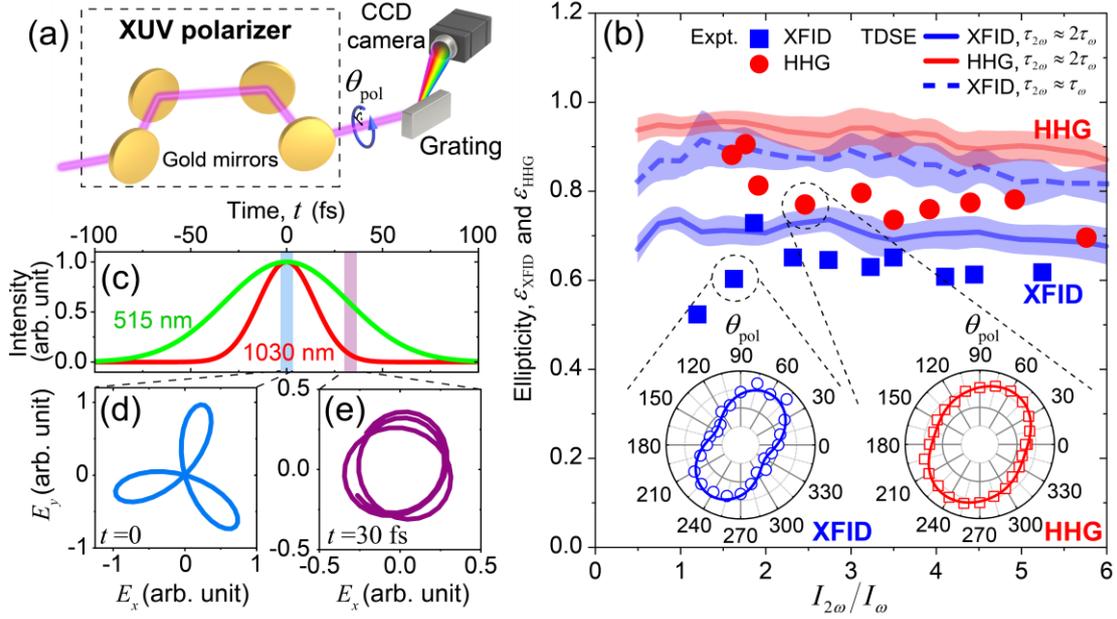

**Figure 2. (a)** Schematic of the XUV polarizer. **(b)** Ellipticity ($\varepsilon$) of the HHG and XFID radiations as a function of intensity ratio ($I_{2\omega}/I_\omega$) from experiments (Expr., symbols) and TDSE simulations (solid and dashed lines). The solid lines represent the TDSE results with the 515-nm pulse duration longer than the 1030-nm pulse ($\tau_{2\omega} \approx 2\tau_\omega$) for HHG (red) and XFID (blue). The blue dashed line represents the TDSE results with equal pulse durations ($\tau_{2\omega} = \tau_\omega$). The shaded areas indicate variations in the simulation results due to total-intensity uncertainty. **Insets:** Typical experimental polarization analysis of XFID (left) and HHG (right). **(c)** Illustration of temporal envelopes of the 515-nm (75 fs FWHM) and 1030-nm pulses (34 fs FWHM). **(d)** and **(e)** Parametric plots of $E_x$ and $E_y$ of the driving laser field within a single cycle of the 1030-nm field at the pulse center ($t$=0) and trailing edge ($t$=30 fs), respectively.

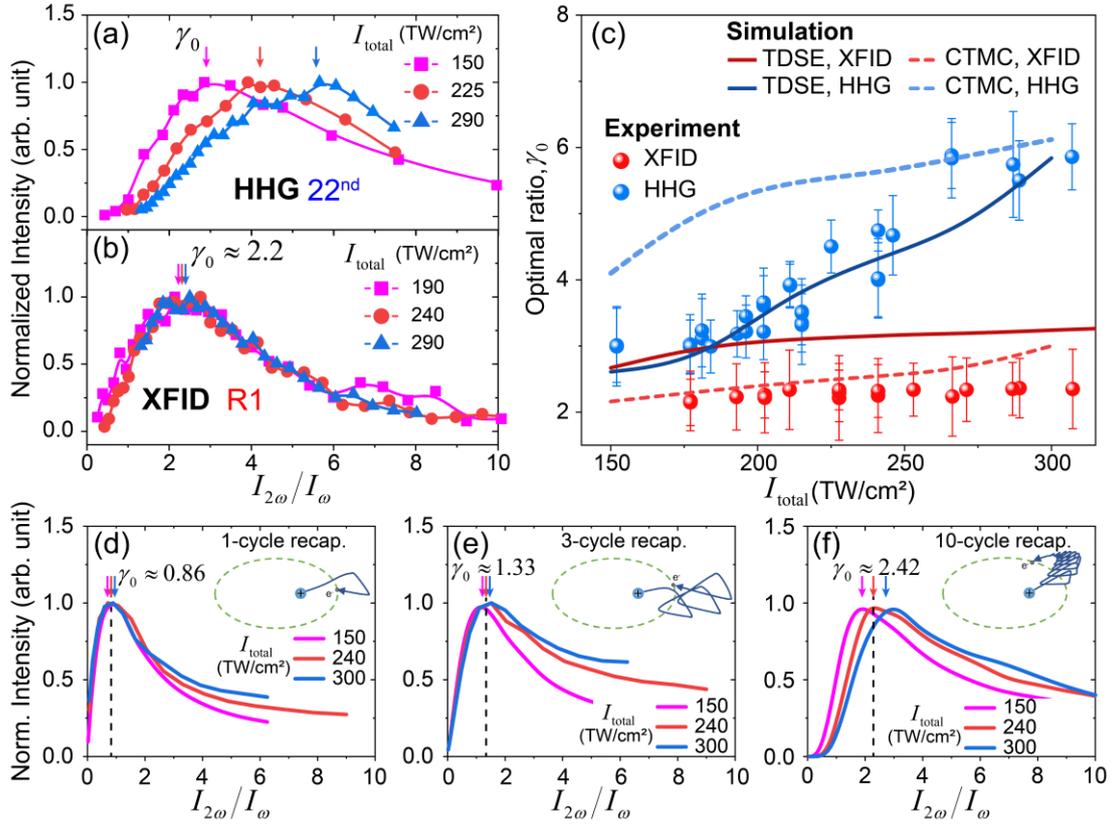

**Figure 3. (a)** and **(b)** Intensity variations of the HHG and XFID radiations as a function of $I_{2\omega}/I_\omega$ under different total intensities ($I_{total}$), respectively. The optimal intensity ratios ($\gamma_0$) are indicated. **(c)** $\gamma_0$ versus $I_{total}$ for XFID and HHG. The symbols represent experimental data, while the solid and dashed lines represent TDSE and CTMC simulations, respectively. **(d-f)** CTMC simulations of the XFID intensity dependence on the intensity ratios $I_{2\omega}/I_\omega$ under scenarios where electrons are recaptured (recap.) in 1 cycle (*N*=0), 3 cycles (*N*=2) and 10 cycles (*N*=9). The corresponding optimal intensity ratios ($\gamma_0$) are indicated. Insets illustrate the respective physical scenarios.

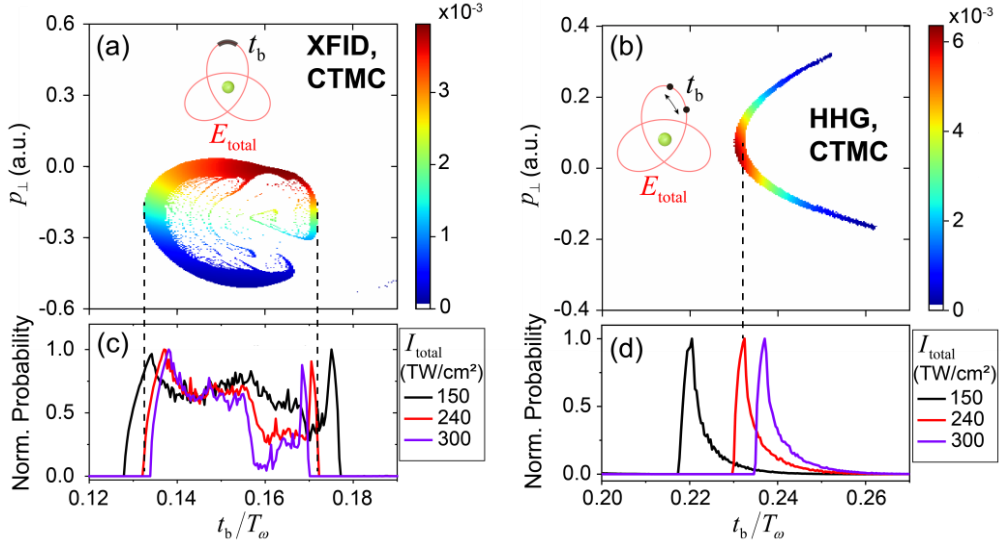

**Figure 4.** **(a)** and **(b)** Electron yields for XFID and HHG, respectively, in the initial ionization coordinates ($t_b$, $p_\perp$) from the CTMC simulations. For HHG, recolliding electrons with energies within a 1 eV interval around the center energy corresponding to the 22nd-order HHG (26.5 eV) are considered. $T_\omega$ represents one optical cycle of the 1030-nm field. a.u. denotes atomic unit. **Insets:** Illustration of the variations in the electron birth time ($t_b$) with respect to $I_{total}$. **(c)** Integrated FTI electron yields versus of $t_b/T_\omega$ for different $I_{total}$. **(d)** Integrated HHG electron yields versus $t_b/T_\omega$ for different $I_{total}$.

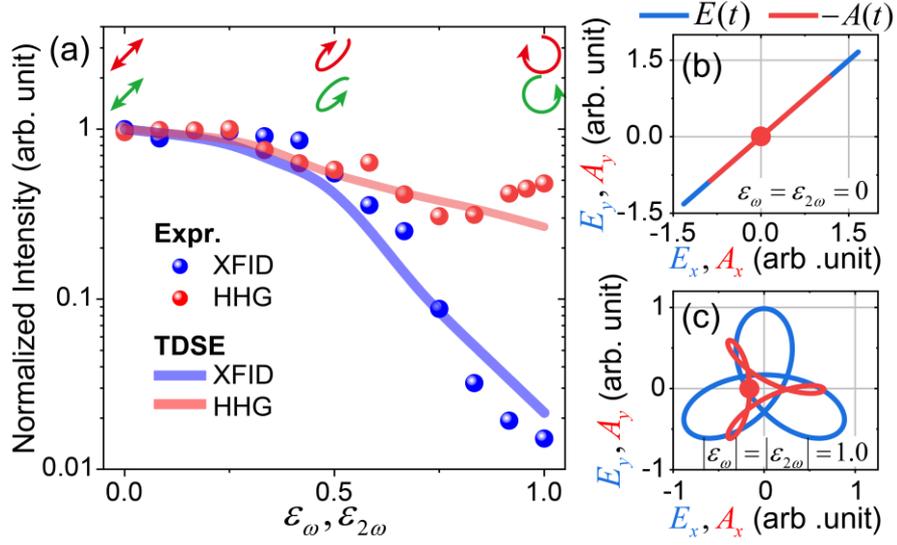

**Figure 5. (a)** Normalized intensities of XFID and HHG radiations as a function of the driving-laser ellipticities ($\varepsilon_\omega$ and $\varepsilon_{2\omega}$). Solid lines represent the TDSE results. The polarization states of the two driving laser fields are labeled. **(b)** and **(c)** Parametric plots of $E_x$-$E_y$ and $A_x$-$A_y$ for the linearly polarized ($\varepsilon_\omega = \varepsilon_{2\omega} = 0.0$) and circularly polarized ($|\varepsilon_\omega| = |\varepsilon_{2\omega}| = 1.0$) driving fields, respectively. The values of -$A(t)$ after the pulse is terminated are labeled by the red dots.